# Examining Platformization in Cultural Production: A Comparative Computational Analysis of Hit Songs on TikTok and Spotify

Na Ta[1 2], Fang Jiao[3], Cong Lin[4], Cuihua Shen[5 *]

[1] Research Center of Journalism and Social Development, Renmin University of China. Email: tanayun@ruc.edu.cn. ORCID: https://orcid.org/0000-0002-6644-3467

[2] School of Journalism and Communication, Renmin University of China.

[3] School of Journalism and Communication, Chinese University of Hong Kong. Email: 1155149759@link.cuhk.edu.hk. ORCID: https://orcid.org/0000-0003-3887-2229

[4] School of Journalism and Communication, Tsinghua University. Email: lin-c23@mails.tsinghua.edu.cn. ORCID: https://orcid.org/0009-0000-8546-8021

[5] University of California, Daivs, USA. Email: cuishen@ucdavis.edu. ORCID: https://orcid.org/0000-0003-1645-8211

[*] Corresponding Author: Cuihua Shen, University of California, Davis, One Shields Avenue, Davis, CA 95616, USA. Email: cuishen@ucdavis.edu

**Abstract**

The (re)creation and distribution of cultural products such as music are increasingly shaped by digital platforms. This study explores how TikTok and Spotify, situated in different governance and user contexts, could influence digital music production and reception within each platform and between each other. Focusing on daily hit song charts as the embodiment of platformization, we collected and analyzed a two-year longitudinal dataset on TikTok and Spotify. We tested the relationships between elements of platformization and hit song popularity within each platform, and examined cross-platform influence flow. Results reveal significant differences in major label, genre, and content features among hit songs on TikTok and Spotify, which can be explained by their distinct platformization practices. We also found some evidence that hit song popularity on Spotify might precede that on TikTok. This study illustrates both the platform-specific mechanisms of TikTok and Spotify and their interconnectedness in the cultural production ecosystem.

# Examining Platformization in Cultural Production: A Comparative Computational Analysis of Hit Songs on TikTok and Spotify

The (re)creation and distribution of cultural products are increasingly shaped by digital platforms. Beyond connecting content creators to a broader group of consumers, digital platforms are also moderating the content of cultural production (Gerrard & Thornham, 2020), configuring the power relations between actors in multi-sided markets (Nieborg & Poell, 2018), and building data infrastructure for the production, distribution, and circulation of cultural products (Siciliano, 2022). As a result, cultural production is increasingly dependent on the cultural, infrastructural, and commercial activities of platforms. This process has been conceptualized as platformization, “the penetration of economic, governmental, and infrastructural extensions of digital platforms into the web and app ecosystems, fundamentally affecting the operations of the cultural industries” (Nieborg & Poell, 2018, p.4276). In the field of digital music, two major players, TikTok with 1.7 billion monthly active users (Iqbal, 2023) and Spotify with 574 million monthly active users (Spotify, 2023), are critically shaping music production and end users’ music consumption.

In the process of platformization in culture production, multiple participants such as content creators, users, and companies are closely involved through platforms more than ever. User-generated data such as content, behaviors (e.g., clicks and subscriptions) and even attitudes (e.g., reactions) are logged and fed into platform algorithms, which iteratively influence the process of culture production. To understand how elements of platformization interact with cultural production, previous studies have focused on platforms’ business model and technical dimension (Nieborg & Helmond, 2019), emerging creative class and algorithmic power (Lin & de Kloet, 2019), and platform governance (van Dijck, 2021), primarily through observational and qualitative approaches.

However, as most existing studies tend to focus on single, rather than multiple, digital platforms, it remains unclear which platformization differences are consequential in shaping the cultural production process, and through what mechanisms. As platforms are owned by different commercial entities, situated in different governance structures, and employ different algorithmic priorities to serve different user populations, it is critical to systematically examine cultural production across *multiple* platforms simultaneously. Moreover, the few multi-platform studies to date (e.g., Hase et al., 2022) tend to focus on comparing and contrasting platformization practices within each individual platform instead of examining the flow of influence across multiple platforms as parts of an ecosystem. This gap limits our understanding of the potential interactions *between* platforms on cultural production, and understanding of platformization as a dynamic and gradual reorganizing process with platforms' participants occupying different positions in the field (Seibt, 2024).

To fill the gap, this study aims to systematically examine platformization and the potential cross-platform flow of influence between two major platforms, TikTok and Spotify, through daily hit songs charts, which represent both aggregated user feedback and algorithmically-curated manifestations of platform influence. By publishing hit song charts, platforms declare (and decide) what is (and what should be) visible and dominant music products. TikTok and Spotify are similar in that their hit song charts are based on the data infrastructure that aggregates feedback from various actors in the multi-sided market, including users and music companies (Toscher, 2021; Aguiar & Waldfogel, 2018). Yet, operationally, TikTok emphasizes users' consumption and re-creation of clipped music products through their fusion with short-form videos, while Spotify is a major player in the distribution and circulation of full-length music products. Based on a two-year longitudinal analysis of Top-100 hit song charts from TikTok and Spotify, this study aims to uncover 1) how these two platforms differ from each other in the processes and mechanisms of platformization and 2) whether there are interactions of music popularity between these two platforms.

The current work makes three contributions to the platformization of cultural production literature. First, the comparative analysis of hit song charts as an embodiment of platformization highlights the platform-specific dynamics of TikTok and Spotify in terms of their distinct strategic priorities in the multi-sided music market (user-oriented vs. producer-oriented, especially major labels), roles in cultural production (creative cultural production vs. content distribution and consumption), and forms of music content (short-form videos vs. full-length music). Second, our cross-platform influence analysis is among the first to find empirical evidence of the interconnected nature of digital platforms. Third, methodologically, this study is among the first to leverage longitudinal computational analysis across two platforms, providing a much-needed empirical complement to the qualitative and cross-sectional analysis found in previous research.

## Theoretical Background

### The Platformization of Cultural Production

The concept of platformization reveals how digital platforms, built on information and communication technologies such as algorithmic moderation and recommendation (Duffy & Meisner, 2023), are becoming dominant and influential infrastructural and economic actors (Helmond, 2015). This process is composed of two elements: 1) the expansion of platforms' data infrastructures (Helmond, 2015), and 2) the expansion of their business models (Poell et al., 2019) to external social webs and communities. Through the process of platformization, the chain of cultural production, including production, circulation, distribution, and consumption, where each process used to happen individually in separate spaces and time points, is recombined within the same uniform space governed by platforms (Morris, 2020).

The first element of platformization involves utilizing the platforms' technical features to connect to other websites and applications, and collecting data to participate in cultural production. The technical features of programmability and modularity facilitated by Application Programming Interfaces (APIs) (Plantin et al., 2018) constitute the technological architecture of platformization. These features allow the functions of platforms to be easily inserted into other websites or applications, which contributes to the connections between platforms and external agents, such as third-party developers (Hind et al., 2022), and construct application ecosystems (Plantin et al., 2018). The platformization process built on these technical features helps the platforms occupy central positions in the markets (Seibt, 2024). In addition to the programmability and modularity of platforms, digital platforms shift from data companies to data infrastructure (Poell et al., 2019), as they connect and integrate data from both the platforms' internal and external databases and make them algorithm- and platform-ready (Helmond, 2015). In music production, utilizing the aforementioned technical features allows platforms to predict and react to popular trends in cultural production, modify cultural producers' behaviors, and further influence musical cultures (Morris, 2020).

The second element of platformization refers to interacting with actors in multi-sided markets (Poell et al., 2019), resulting in different business models as platforms tackle different relationships and priorities. The typical multi-sided market in cultural production involves end-users, cultural producers, institutions, and advertisers (Nieborg et al., 2022), the interactions among whom jointly shape the platforms as aggregators of institutional and economic connections (Poell et al., 2021). The platforms would implement different business models and strategies due to their unique positions in the multi-sided market. For example, Netflix mainly profits from subscriptions rather than from advertisements. Therefore, Netflix needs to deal with content creators in the media industry, such as movie studios and television distributors, as well as subscribers (Nieborg et al., 2022). Meanwhile, YouTube emphasizes social interaction and e-commerce in its business model, hence YouTube needs to focus on more shifting relationships between users, producers, advertisers, and institutions (Ørmen & Gregersen, 2023). In music cultural production, the multi-sided nature of the digital music market not only includes users, producers, and advertisers but also different types of platforms that have various platformization strategies to fulfill their needs of monetization and cater to users' preferences (Seibt, 2024). As such, a comparative analysis of how multiple platforms in music production differentiate from and influence each other is necessary to understand platformization as a gradual reorganizing process (Seibt, 2024).

**TikTok and Spotify**

This study compares user-generated short video content platforms and music streaming platforms, represented by TikTok and Spotify, respectively, to understand the dynamics and effects of platformization within an ecosystem. We identify three main differences between these two types of platforms. First, they prioritize distinct actors in the multi-sided music market, where TikTok is user-centric and Spotify is producer-centric. Second, they play different roles in music cultural production, with TikTok highlighting creative cultural production and Spotify emphasizing content distribution and consumption. Third, music is presented as a backdrop in short-form videos on TikTok, and as full-length music products with lyrics and metadata on Spotify.

User-generated short video platforms, such as TikTok and Douyin, prioritize creative interaction over discursive interaction (Zulli & Zulli, 2022). These platforms actively participate in the production of music by providing clips of popular songs as background music for videos (Wang & Wu, 2021). Furthermore, the platforms algorithmically manipulate the visibility of user-generated videos to moderate cultural production activities (Duffy & Meisner, 2023). Therefore, platforms participate in cultural production through governance and shape an “experience economy” as a result (Zhang & Negus, 2021, p. 539). As such, users become active creators using music rather than passive listeners on music streaming platforms. By contrast, music streaming platforms, such as Spotify and Apple Music, focus on delivering licensed music content in full versions to users through telecommunication networks (Morris & Powers, 2015). On music streaming platforms, music is produced by professionals, and the platforms act as distributors or curators instead of producers (Prey et al., 2022). Compared to user-generated short video platforms, user engagement on streaming platforms is limited to creating playlists, yet user-generated playlists are marginal compared to platform-generated playlists (Bonini & Gandini, 2019). Through playlists that demonstrate the platforms’ curatorial power, music streaming platforms expand their influence on music production, advertisement, and even the financial market (Prey et al., 2022).

To examine music production on the two platforms, we consider hit song charts as an embodiment of platformization. Since 1984, Billboard has provided popular song charts based on album sales (Chart Data, n.d.). With the emergence of digital music platforms, Billboard charts have started to include trends data on these platforms (Aum et al., 2023). Meanwhile, digital platforms also release their own hit song charts, which are based on data contributed by various actors in the multi-sided market, such as streaming and subscription data from users, advertising data, and promotion campaigns of music companies (Toscher, 2021; Aguiar & Waldfogel, 2018). Such data are then processed through proprietary and opaque algorithms that assign different weights according to platforms’ priorities (Aguiar & Waldfogel, 2018; Spotify, n.d.). Thus, hit song charts embody the processes and elements of platformization. As the most visible indicator of the music goods success, hit song charts influence cultural production on platforms (Aguiar & Waldfogel, 2018). Comparing and contrasting hit song charts between TikTok and Spotify thus offers a systematic examination of platformization dynamics within each platform and their possible interaction with each other. Below, we discuss the potential drivers for hit song popularity, including record label, music theme and genre, then discuss how popularity on one platform might potentially spill over to another.

**What Contributes to Hit Song Popularity?**

Record labels that produce and sell music could influence whether and to what extent songs are popular. Major labels include Sony Music Entertainment, Universal Music Group, and Warner Music Group (Marshall, 2013). Smaller enterprises that operate outside of the above integrated and well-financed corporations are referred to as independent labels (Marshall, 2013). This split reflects disparities in the resources available to companies to produce and promote music (Ren & Kauffman, 2017) as well as target audience distinctions. Independent labels have lower operating costs and more localized business models than major labels, occupying niche markets both geographically and culturally (Hesmondhalgh & Meier, 2015). Therefore, their music might be less popular globally when compared to songs produced by major labels, which command greater resources and target a global audience.

Traditionally, album sales are considered a hallmark of music popularity. Major labels' abundant promotional resources make their songs easier to reach target listeners and therefore a higher likelihood of potential sales, compared to songs from independent labels (Ren & Kauffman, 2017). Meanwhile, independent labels lack efficient channels or enough resources to promote their songs, resulting in less popularity. On music streaming platforms, music popularity is related to not only album sales, but also users' listening and sharing data processed through proprietary algorithms (Spotify, n.d.). It is not clear whether major labels could maintain their dominant position, as independent labels may gain more exposure opportunities on streaming platforms compared to those of traditional music markets. For example, the collaboration between Spotify and independent artists increases exposure for the songs they produce and helps elevate their popularity around the world (Prey et al., 2022).

Spotify tends to promote song lists that are created and curated by the platform and contain a mixture of songs produced by major labels and independent labels rather than promote the playlists curated by major labels only (Prey et al., 2022). This enhances the platform's curatorial freedom in deciding which songs, regardless of label, can be promoted and weakens major labels' impact on song popularity. TikTok is often used as a marketing platform before the official release of music (Coulter, 2022). Major labels' investment in marketing on TikTok could improve the visibility of their songs, which contributes to the songs' popularity. Simultaneously, the low cost of creating short videos allows independent musicians to use short video platforms to increase their songs' popularity. We are interested in whether the distinction between major labels and independent labels on digital platforms still correlates with music popularity, and whether the different positions and strategies of TikTok and Spotify adopt to interact with actors in the multi-sided music markets would translate into different shares of major and independent labels. Therefore, we ask,

*RQ1*: (a) Is the popularity of a hit song associated with its record label? (b) Are there significant differences in hit songs' record labels on TikTok and Spotify?

Themes could also contribute to music popularity. Traditionally, relationships and romance are prominent themes in popular music (Christenson et al., 2019). These themes continue to dominate on streaming platforms (Ren & Kauffman, 2017). Other themes that appear in hit song charts include dance, anger, violence, and drugs (Berger & Packard, 2018; Yu et al., 2023). On music streaming platforms, hit songs are presented in full-length versions, and themes can be identified through the analysis of lyrics (Yu et al., 2023). By contrast, existing studies that examine popular themes on TikTok focus more on video content and hashtags (Zhang, 2021), such as #GlobalWarming, #ClimateChange, and #ForClimate hashtags on various social or political issues (Hautea et al., 2021; Zeng & Abidin, 2021). Furthermore, as TikTok promotes music challenges to encourage interactions among users, themes that mobilize users' sharing and interactive behaviors may be more popular on TikTok. Therefore, we ask,

*RQ2*: (a) Is the popularity of a hit song associated with its theme? (b) Are there significant differences in hit songs' themes on TikTok and Spotify?

Music genres may also influence music popularity, and its distribution may vary across platforms. Traditionally, genres such as Electronic, Funk/Soul, and Rock, are found to be associated with success on Billboard (Askin & Mauskapf, 2017). Music genres arise from the cultural and social background of the music production. For example, Hip-hop as a music genre is produced as a universal means of resistance in the face of marginalization (Krogh, 2017). The emergence of new genres is related to the creation of musical-cultural communities and scenes (Krogh, 2023). The platformization process provides the necessary data infrastructure to support diversified cultural communities (Poell et al., 2019), thus promoting the emergence of diversified music genres (Petrovic, 2023). Meanwhile, this process also obfuscates the relationship between genres and music popularity. On TikTok, music genres suitable for "music challenges" could become prominent, such as Pop, Rock, and Dance. Users participating in these challenges are grouped together by hashtags - they create videos using the same audio resources, and invite other users to participate in the challenges. They range from dance challenges, singing challenges to lip-syncing challenges (Vizcaíno-Verdú & Abidin, 2022; Petrovic, 2023). By contrast, Spotify's popular songs show higher genre diversity and the blurring of boundaries between genres, as its recommendation mechanism focuses on personalization and the platform's computerized genre analysis produces an infinite array of genre categories (Krogh, 2023). Therefore, we ask,

*RQ3*: (a) Is the popularity of a hit song associated with its genre? (b) Are there significant differences in hit song genres on TikTok and Spotify?

**Cross-Platform Interactions of Music Popularity**

Platformization increases the integration of web applications and the flows of data (e.g., posts, music, and user information) since platforms not only collect data of their own but also expand collection techniques to external apps (Helmond, 2015). This cross-platform circulation of digital data reflects the interconnected nature of platforms, yet the their interactions have rarely been examined. Recent literature emphasizes that platformization is a dynamic process through which platforms and the organizations behind them compete to establish themselves in a specific field or industry (Plantin et al., 2018; Seibt, 2024). Therefore, platformization processes and dynamics cannot be fully understood by focusing on an individual platform in isolation. Instead, such analysis must account for the conditions of wider social contexts as platforms engage with various actors within multi-sided markets (e.g., consumers, trade associations, or advertisers) to solidify their power positions (Seibt, 2024). In the music industry, both TikTok and Spotify are key players striving to maintain and grow their markets. As an embodiment of platformization, the hit song charts on TikTok and Spotify thus may offer a glimpse of the cross-platform interplay between users, content creators, and algorithms that co-exist in the multi-platform ecosystem.

First, users are the ultimate engine that drives the monetization success of digital platforms. To attract and retain users' attention, platforms are motivated to adjust their interfaces and functions in response to users' practices and needs (Poell et al., 2019). Therefore, as users hop between platforms to chase the most viral content, platforms are also motivated to optimize their content offerings to engage a large user base. Prior empirical studies have suggested that online viral contents have spillover effects across platforms (i.e., the content popularity could flow from one platform to another; Krijestorac et al., 2020) while users using multiple platforms could act as bridges for cross-platform information flow (Jiang et al., 2016). Specifically, Spotify is an established music distribution channel (Vonderau, 2019), whereas TikTok is known as a playful community around dances and music, liked by half of the US population (Cervi & Divon, 2023). Their shared user base could potentially act as pollinators for music popularity interplay across platforms.

Second, content creators are motivated to maximize the exposure of their cultural products on not one, but many platforms (Glatt, 2021). Further, instead of publishing the exact same content on multiple platforms, they typically adapt to each specific platform based on their understanding of their affordances and algorithmic priorities (Arriagada & Ibáñez, 2020). For example, artists on Spotify are motivated to "optimize" their music by altering its style in an "attention-getting" way (Morries, 2020). On TikTok, content creators could add background music to their videos and leverage affordances such as dance challenges, lip-syncs, and trending hashtags to increase engagement and exposure (Toscher, 2021; Zeng & Kaye, 2022). Research showed that TikTok videos are more likely to be found, liked, and shared if the sound effects are replicated from and linked to an already popular video (Zulli & Zulli, 2020). Therefore, content creators looking to maximize their content exposure across multiple platforms may catalyze the cross-platform spillover of music products.

Lastly, algorithms may facilitate cross-platform influences.  Both TikTok and Spotify have implemented algorithms for content curation. With a virality-centric algorithmic logic, TikTok motivates users to create viral content to compete for visibility (Zeng & Kaye, 2022). On Spotify, the music playlists are determined jointly by editorial tastemakers and a suite of proprietary machine-learning algorithms (Bonini & Gandini, 2019). With these algorithms, two underlying mechanisms could lead to cross-platform interactions between TikTok and Spotify. First, the algorithmic-driven exposure of a hit song on one platform may enhance users' attention and awareness, which in turn contributes to more searches on another platform (Liang et al., 2019). Second, input data for algorithms on one platform may come from external applications through APIs (Helmond, 2015). For example, music trending data from other platforms may feed into TikTok algorithms so as to fully capture multi-platform music buzz. In sum, multiple actors potentially pave the way for cross-platform influences of hit songs between TikTok and Spotify. We thus propose the following research question:

*RQ4*: Does the popularity of hit songs on TikTok and Spotify influence each other?

## Methods

### Data Collection and Processing

Data were collected through Chartmetric, a music analytics service that provides hit song data from popular music platforms. We collected the top 100 hit songs from TikTok and Spotify on their daily charts between June 1, 2020 and May 31, 2022. Each hit song record includes information of song name, duration, rank, album name, and artists/composers.

Data cleaning and processing were performed in two steps. First, due to missing and erroneous data, we removed or supplemented a few records to ensure the quality of analysis (see Appendix A for details). After merging the song-day charts on each platform and removed duplicates based on song names and unique IDs, there were 348 distinctive hit songs on TikTok and 1,707 on Spotify. Certain hit songs have incomplete information, such as record label and duration, so we manually complemented the dataset by locating such information on Wikipedia. We removed 27 TikTok songs for which duration information could not be found. The final dataset before coding contained 321 TikTok hit songs and 1,707 Spotify hit songs (2,028 songs in total, with 68 appearing on both platforms).

Second, we located genre and lyrics information from additional sources. The genre of hit songs on Spotify was available from Chartmetric. For TikTok, we manually collected the genre information from Wikipedia and Music Genre Finder function of Chosis[1] (if Wikipedia page was unavailable). The genre information was later transformed into a 5-category genre measure (see Measures). Lyrics were obtained using the API of Genius[2], supplemented by Google Search if not unavailable on Genius. For hit songs on TikTok, we obtained lyrics of the full song even though most TikTok versions are snippets of the original version, in order to avoid incomplete semantics. Some songs are cover versions (or adaptations) of an original song, typically with the same lyrics but different singers or record labels. We then created a Group ID variable, and manually grouped the cover versions with their original song through the same Group ID. Table 1 shows the descriptive statistics of the dataset.

## Measures

### *Music Popularity*

Music popularity was measured by two indicators. First, on_chart_days of a song refers to the total number of days the song appeared in the daily Top-100 charts between June 1, 2020, and May 31, 2022 ($M_{TikTok}$ = 208.816; $M_{Spotify}$ = 42.531). It was log transformed before analysis due to skewness. Second, daily popularity (1 to 100) was obtained by subtracting the raw ranking value obtained from Chartmetric from 101, so that a larger value reflects higher popularity. If a song was off the Top-100 chart on a certain day, its daily popularity was '0'. Therefore, on_chart_days is a cumulative popularity measure for each individual song, while daily ranking is a time-varying popularity measure.

### *Major label*

---

[1] Chosis is a tool for searching and discovering music information, https://www.chosic.com/music-genre-finder/

[2] Genius is an online collection of lyrics and music knowledge, www.genius.com

Major label refers to any record label owned or controlled by Universal Music Group, Sony Music Entertainment, or Warner Music Group (Marshall, 2013), and record labels that operate outside of these three companies are considered independent labels. We created a codebook that included all major record labels controlled by the above three music groups according to their official websites.[3] The label of each hit song was coded as '1' if it was produced by major labels in our codebook; or '0' otherwise.

***Genre***

Each Spotify song was categorized into at least one genre among the 344 categories that Chartmetric provided. For TikTok songs, we manually complemented their genre information from Wikipedia, or Chosis if the songs do not have Wikipedia entries. However, the final set of genre was too dispersed to conduct statistical analysis. Therefore, we conducted a word frequency analysis of genres of hit songs in our dataset to identify the main music genres, based on the genres returned by Chartmetric and manual collection. The results showed that the most frequent words were Pop (25.7%), followed by Rap (16.5%), Hip-hop (11.5%), Trap (4.98%), Dance (2.6%), and R&B/Soul (2.1%). Drawing on previous literature (Choi & Downie, 2019), we derived five music genres: Pop, Rock, Hip-hop/Rap/Trap, R&B/Soul, and Dance. We added the dance category because dance is prominent in cultural activities on TikTok (Zulli & Zulli, 2022) and is also among the top five genre categories in the word frequency analysis. We then labeled each song's genre accordingly, if its genre information provided by Chartmetric or manual supplementation contains one of the above five categories keywords. As such, a song can be categorized into multiple genres. If the genre information in the dataset did not contain any of the above five categories, it was classified as 'others' ($N$ = 331, 16.321%; 61 in TikTok and 270 in Spotify). For example, if the genre information provided by Chartmetric or supplemented manually includes the word 'soul', it would be labeled with 'R&B/soul'. Labeling of genres was not mutually exclusive, each song could have one or more categories of genres.

[3] Universal Music Group: https://www.universalmusic.com/labels/; Sony Music Group: https://www.sonymusic.com/labels/; Warner Music Group: https://www.warnerrecords.com.

***Theme***

We conducted a content analysis on hit song lyrics. Based on a random sample of about 11% of hit songs ($N$ = 224) in our dataset as well as prominent themes found in previous studies (Christenson et al., 2019; Berger & Packard, 2018), we identified five themes in the lyrics: Love/Passion/Relationship/Desire, Festival/Partying/Good times, Identity/Depression/Mood Issues/Spirituality, Social/Politics/Violence/Race/Religion, and Others (see Table S2 for descriptions). Three graduate students majoring in communication coded the themes of hit songs. Themes were not mutually exclusive, and each song could express more than one theme. Instrumental music without lyrics was coded as missing value ($N$ = 10, 0.49%). Hit songs that could not be classified into any of the above themes were coded as 'Others' ($N$ = 115, 44 in TikTok and 71 in Spotify). The composite Cohen's kappa reliability was satisfactory (see Table S2).

***Control variables***

Control variables included the duration of each song (in minutes), their peak daily popularity, and the singer's popularity. For peak daily popularity, we identified the highest daily rank of each hit song between June 1, 2020, and May 31, 2022, and subtracted it from 101. Singer popularity was measured by the number of followers of the singer (or the artist, for instrumental music) as of February 2023, from each singer's page on Spotify, since Chartmetric API did not provide this information. For singers without personal pages or follower statistics, their popularity was coded as '0'. We then log transformed singer popularity due to skewness. Group ID identified if some songs were different versions of the same original song, and it was included as a random effect to control for song-level variance.

**Analysis**

We conducted a three-part analysis. First, we estimated two separate mixed-effects linear regression models, with Group ID as a random intercept, to test the associations between major label (RQ1a), theme (RQ2a), and genre (RQ3a) and the popularity of hit songs on TikTok and Spotify, respectively. The dependent variable is music popularity, measured by on_chart_days of the song, and control variables included song duration, peak daily popularity and singer popularity. Second, the differences between TikTok and Spotify (RQ1b, RQ2b, and RQ3b) were tested using independent samples *t*-tests (and paired sample *t*-tests, see Appendix B) of the daily percentage of major labels, genres and themes of the top 100 hit songs (*N*=721 on TikTok and *N* = 726 on Spotify).

Third, to answer RQ4, we first focused on those hit songs that appeared at least once, but not necessarily simultaneously, on both platforms' Top-100 charts between June 1, 2020 and May 31, 2022, and we analyzed the time sequence order of the hit songs' appearance on both platforms. To maximize the sample size, we considered cover versions and the original song together, based on their Group ID (*N* = 68). Then we used vector autoregression (VAR) models to investigate the mutual influence of hit songs' popularity on two platforms, as VAR analysis can assess what impacts variables have on each other over time (Brandt & Williams, 2007). For VAR analysis, we focused on a sample of hit songs that appeared at least 2 times on both TikTok and Spotify independently, with at least 1 day in which the hit song appeared on both platforms simultaneously (*N* = 30), and then created their weekly ranking time series (see Appendix C for details of time series construction). We conducted the Augmented Dickey-Fuller (ADF) tests on the time series, and calculated the difference between consecutive time points to fulfill the stationarity assumption of the VAR model (Box-Steffensmeier et al., 2014). We determined the appropriate number of lags for each VAR model based on Akaike's Information Criterion (AIC). And then we conducted Granger causality tests and leveraged the cumulative impulse response functions (CIRFs) and the forecast error vector decomposition (FEVD) to assess the direction and the size of the influences (see Appendix D for details of VAR analysis).

## Results

### Comparison of TikTok and Spotify

RQ1a, RQ2a, and RQ3a concern the relationship between record labels, theme, genre and hit songs' popularity (Table 2). Compared to hit songs produced by independent labels, songs produced by major labels were less popular on TikTok ($B$ = -.628, $SE$ = .234, $p$ = .007) and on Spotify ($B$ = -.221, $SE$ = .079, $p$ = .005). When examining the relationship between themes and the popularity of hit songs on both platforms (RQ2a), none of the themes is significantly associated with the popularity of hit songs on TikTok. On Spotify, Relationship/Love/Passion/Sexual Desire was positively associated with song popularity ($B$ = .342, $SE$ = .085, $p$ < .001), while Social Politics/Anger/Violence/Race/Religion was negatively associated with popularity ($B$ = -.372, $SE$ = .102, $p$ < .001). In terms of genre, dance is significantly associated with hit songs popularity on TikTok ($B$ = -.926, $SE$ = .301, $p$ = .016), but none of the genres is significantly related to Spotify's hit song popularity.

RQ1b, RQ2b, and RQ3b concern the differences between TikTok and Spotify in the distribution of major labels, genres, and themes among hit songs (Figure 1). Independent samples $t$-tests of daily percentages reveal that compared to TikTok, Spotify has a higher proportion of songs produced by major labels ($M_{TikTok}$ = 50.277; $M_{Spotify}$ = 75.602, $t$ =112.37, $df$ = 1445, $p$ < .001). In terms of genre, compared to TikTok, Spotify contains a lower proportion of R&B/Soul ($M_{TikTok}$ = 13.605, $M_{Spotify}$ = 12.030, $t$ = -11.146, $df$ = 1445, $p$ < . 001) and Dance ($M_{TikTok}$ = 24.782, $M_{Spotify}$ = 19.792, $t$ = -32.012, $df$ = 1445, $p$ < . 001), but a higher proportion of songs in Pop ($M_{TikTok}$ = 42.730, $M_{Spotify}$ = 69.118, $t$ = 184.55, $df$ = 1445, $p$ < . 001) and Hip-hop/Rap/Trap ( $M_{TikTok}$ = 33.588, $M_{Spotify}$ = 39.809, $t$ = 20.168, $df$ = 1445, $p$ $< .001$). In terms of themes, compared to TikTok, Spotify has a lower proportion of songs on Good Times/Partying/Carnival/Festival ($M_{TikTok} = 7.38$, $M_{Spotify} = 5.963$, $= -5.135$, $df = 1445$, $p < .001$) and Social Politics/Anger/Violence/Race/Religion ($M_{TikTok} = 14.770$, $M_{Spotify} = 10.634$, $t = -22.542$, $df = 1445$, $p < .001$), but a higher proportion of songs about Relationship/Love/Passion/Sexual Desire ($M_{TikTok} = 54.290$, $M_{Spotify} = 75.387$, $t = 84.846$, $df = 1445$, $p < .001$) and Identity/Depression/Mood Issues/Spirituality ($M_{TikTok} = 15.325$, $M_{Spotify} = 28.798$, $t = 61.003$, $df = 1445$, $p < .001$).

**Cross-Platform Interactions of Music Popularity**

RQ4 asked whether the popularity of hit songs on TikTok and Spotify influenced each other over time. We first examined the temporal order in which the hit songs appeared on the TikTok and Spotify daily Top-100 charts. 68 hit songs appeared at least once on both platforms, with or without overlapping day(s), and 38 songs appeared at least twice on both platforms (see Figure 2 for their daily appearance sequences). A visual inspection shows that most hit songs entered and exited first on Spotify's Top-100 daily charts before they did on TikTok charts. Chi-square tests confirmed this observation. Among the 68 hit songs appeared on both platforms' Top-100 charts at least once, 50 (73.8%) of them entered the Spotify daily charts earlier than TikTok charts, while only 8 (11.8%) of them entered TikTok daily charts first, $\chi^2(1, N = 68) = 21.086$, $p < 0.001$. Meanwhile, 56 hit songs (82.35%) exited Spotify daily charts before doing so on TikTok, while only 10 exited TikTok daily charts first, $\chi^2(1, N = 68) = 48.271$, $p < 0.001$. This result remained robust when we removed time-censored records (i.e., hit songs that appeared on the first and last days of our observation time window; See Appendix E).

To further explore the cross-platform interactions of music popularity, we created a weekly popularity time series of hit songs that 1) appeared on both platforms at least 2 times independently, and 2) had at least one day of overlap on both platforms (*N* = 30). 30 VAR models were estimated to predict the popularity of each individual week (see Table S5).

When investigating Spotify's influence on TikTok, the Spotify popularity of six songs (*All I Want for Christmas is You, Señorita, You Got It, Rags2Riches, Life Goes On, Sad Girlz Luv Money*) Granger caused their subsequent popularity on TikTok. However, further CIRF results indicated that, for all these six hit songs, the impact of their Spotify popularity on TikTok popularity approached zero (e.g., nonsignificant).

When investigating TikTok's influence on Spotify, we found that TikTok popularity Granger caused Spotify popularity for seven songs, five of which had stable and significant CIRF results (see Table S5). One standard deviation increase in the TikTok popularity of *All I Want for Christmas Is You* (*SD* = 33.015), *SugarCrash* (*SD* = 17.654) and *Sad Girlz Luv Money* (*SD* = 17.856) resulted in a decrease of 7.208, 2.246, 5.034 in their Spotify popularity after 4 weeks, respectively. For *Life Goes On* (*SD* = 14.771) and *Love Nwantiti* (*SD* = 5.929), one standard deviation increase in their TikTok popularity decreased about 5.576 and 12.426 Spotify popularity after 3 weeks, respectively. In other words, as these songs rose in the TikTok charts, their popularity waned on Spotify, again suggesting that their Spotify popularity precedes that of TikTok. These five hit songs were all produced by major labels, but had different genres and themes.

## Discussion and Conclusion

The (re)creation and distribution of cultural production are increasingly dependent on the cultural, infrastructural, and commercial activities of platforms in multiple dimensions. Yet, previous research has not systematically explored how different platformization processes and dynamics are manifested through cultural production, and whether platforms influence each other. To address this gap, our research considers hit song charts as an embodiment of platformization and examines the features of hit songs and their popularity in the music market from a cross-platform perspective. We collected and analyzed a two-year longitudinal dataset based on daily hit song charts of TikTok and Spotify from 2020 to 2022. Our findings reveal that, compared to Spotify, TikTok hit songs were more centralized, stayed on Top-100 charts longer, and were more likely to be produced by independent labels. Our results also provide some indication that Spotify hit song popularity precedes that of TikTok. Taken together, this study reveals that the distinct platformization practices observed on TikTok and Spotify are not only influenced by their strategic priorities in the cultural production process, the market segment they aim to serve, and the preferred forms of music content, but also the platformization strategies of other digital platforms in the same or adjacent cultural space.

**Popularity was More Centralized and Long-lasting on TikTok than on Spotify**

Within the two-year time window, TikTok daily Top-100 charts contained far fewer unique songs ($N$ = 321) than those of Spotify ($N$ = 1707), and these hit songs tended to stay longer on the charts, indicating a greater degree of hit song centralization and longevity on TikTok. In other words, there are significantly fewer hit songs dominating the TikTok Top-100 charts, and their popularity was stickier than their counterparts on Spotify. This phenomenon can be a result of TikTok's algorithmic orientation to accentuate already viral content. TikTok algorithms recommend already popular songs to users, which in turn enhances the visibility of popular content (Zeng & Kaye, 2022). The music challenges, as a popular form of creative practice, also contribute to the emergence of the centralized and long-lasting song popularity on TikTok as users in challenges invite more users through hashtags, creating more content with the same audio resources (Vizcaíno-Verdú & Abidin, 2022; Zhang, 2021; Zulli & Zulli, 2022).

Meanwhile, songs produced by major record labels are far more represented in the hit song charts of Spotify than that of TikTok, where major record labels and independent labels were roughly evenly split. This result is consistent with existing research that major record labels remain dominant on digital streaming platforms like Spotify (Marshall, 2013; Mall, 2018), while independent artists enjoy more visibility on user- and community-centric platforms such as TikTok (Sadler, 2022). This highlights that Spotify and TikTok each have distinct platformization strategies in the multi-sided market: Spotify's strategy prioritizes established producers, especially major labels, while TikTok is more aligned with users and creative communities. However, we also found that hit songs produced by major labels tend to be less popular (ranked lower) than those produced by independent labels, and this pattern is observed on both TikTok and Spotify.

**TikTok and Spotify Favor Distinct Themes and Genres**

We also found that select themes are associated with hit song popularity on Spotify, but not on TikTok. Spotify focuses on streaming full-length music and provides detailed metadata, including lyrics (Dhaenens & Burgess, 2019). Such information allows users to fully appreciate the themes and messages of hit songs. In contrast, TikTok is a short-form video platform where clipped snippets, rather than full-length songs, serve as an auxiliary to the video content (Zeng & Abidin, 2021). As a result, lyrics are typically edited into short clips without the full verse. This again highlights the two platforms' different priorities in cultural production: Spotify is a platform primarily focused on *distributing* full-length music products with both auditory and textual information, while TikTok prioritizes creative cultural practices where music is an important, yet far from the only, ingredient. Furthermore, our results are consistent with previous research that songs about Love and Relationships tend to be more positively associated with the popularity of hit songs on Spotify (Ren & Kauffman, 2017). We also found that Social/Politics/Violence/Race/Religion is negatively related to hit song popularity on Spotify.

We found no genre association with music popularity on either TikTok or Spotify, with one exception - Dance was positively associated with song popularity on TikTok. This result might highlight TikTok's emphasis on user engagement and its popular content format "dance challenges" (Zulli & Zulli, 2022; Petrovic, 2023), which promotes users' creative content production based on certain music and dance moves. Meanwhile, with the exception of Rock, the distribution of hit song genres on these two platforms is notably different, with mainstream genres, including Pop and Hip-hop/Rap/Trap, taking more higher on Spotify than they are on TikTok. As such, our findings again confirm the different platformization priorities: TikTok tries to engage users as creative content producers in the multi-sided market, while Spotify treats users as consumers and subscribers by providing mainstream music products.

**Spotify Popularity Might Precede That of TikTok**

Exploring the possible interactions between TikTok and Spotify, our results first indicate that only a small number of hit songs (*N*=68) appeared on both platforms within the two-year observation period, most of which entered and exited the daily Top-100 charts on Spotify before they did on TikTok. Additionally, the VAR analysis of weekly hit song popularity reveals that, for a few songs, all of which were produced by major labels, their popularity on TikTok negatively predicted their subsequent popularity on Spotify. Taken together, our analysis provides limited evidence that hit song popularity on Spotify might precede that of TikTok.

One possible explanation for the observed unidirectional popularity flow lies in the different roles TikTok and Spotify play in the music production industry. As mentioned previously, Spotify attained its success by centering on digital content curation, distribution, and marketing (Vonderau, 2019). Major record labels, such as Universal Music Group, have established processes and better resources to support promotion and marketing campaigns on Spotify than independent labels. Upon release of new songs, major labels could strategically place their tracks on popular Spotify playlists, which helps increase the song's initial exposure with consumers (Morris, 2020). This initial music popularity among users is likely to follow the "success-breeds-success" dynamic (Van de Rijt et al., 2014), as the initial visibility gets amplified in a positive feedback loop, resulting in sustained attention from users and higher rankings on algorithmically generated charts over time.

By contrast, TikTok has emerged as a platform for creative practices, especially among the younger generations (Toscher, 2021). Once a track becomes popular on Spotify or elsewhere, users of both Spotify and TikTok might act as cross-pollinators (Jiang et al., 2016) by re-distributing and re-creating the original song into various TikTok formats through memetic templates such as "lip-syncing" or "duets" (Cervi & Divon, 2023). This process could be further amplified by TikTok's powerful algorithms that prioritize viral and engaging content. For example, TikTok would capitalize on the latest music trends, discovered on its own platform as well as elsewhere, and recommend popular songs to be used in users' content creation process. For example, it may promote sound imitation by suggesting that users contribute to a "sound grouping" in which a group of videos are similar in sound but different in content (Zulli & Zulli, 2020). Meanwhile, content creators on TikTok are incentivized to use catchy hit songs in order to become 'algorithmically recognizable' (Gillespie, 2017, p. 63) since there is a greater chance to be found, liked, and then shared (Zulli & Zulli, 2022). Through these actors and mechanisms, hit songs could extend their popularity on TikTok after gaining visibility on Spotify.

In summary, the observed differences and interconnectedness in hit song charts between TikTok and Spotify can be attributed to their distinct platformization practices. First, in multi-sided markets, TikTok tends to prioritize users' engagement as the platform focuses on promoting viral, catchy music that motivates users' participation and creative cultural production. In contrast, Spotify's business model emphasizes its role as a distribution channel in cultural productions. Second, their influences vary in different stages of music production, with Spotify focusing more on the release and distribution, and TikTok focusing on users and prosumers and content producers. From this perspective, their platformization strategies are the result of competitive rivalry and the promotion of vertical integration in the industry. Third, the two platforms differ in terms of technical affordances. TikTok prioritizes viral short-form video content, using catchy music as backdrops, while Spotify offers full-length songs with detailed information, appealing to a broad range of tastes.

## Limitations

This study has two main limitations that warrant future research. First and foremost, we selected TikTok and Spotify for this comparative study because they both publish daily hit song charts and are the most popular players in the short video and music streaming markets, yet they have important differences in technology, algorithms, genre and users populations. It is therefore difficult to pinpoint with confidence the exact mechanisms and characteristics of these platforms that contributed to the observed difference in our comparison. Second, the data collection period coincided with the first two years of the COVID-19 pandemic, when many countries implemented stay-at-home orders. These changes may have significantly influenced music and video production and consumption patterns, as manifested by the hit song charts. To replicate our analysis, future research is encouraged to conduct more comprehensive comparative studies across more digital platforms and over longer periods.

## Conclusions

Examining hit song charts as embodiment of platformization, this study represents one of the first computational attempts at uncovering platformizaton practices in the digital music space. Overall, we found that hit songs on TikTok and Spotify manifest distinct features in genre, themes and record label. These findings can be attributed to their platformization strategies in the multi-sided music market (user-oriented vs. producer-oriented, especially for major labels), primary roles in cultural production (creative cultural production vs. content distribution and consumption), and forms of music content (short-form videos vs. full-length music). Yet, despite these differences, we also found limited evidence for cross-platform influence, through the joint forces of users, creators, and algorithms with the multi-platform ecosystem.

## Data Availability Statement

Data and code are available at https://osf.io/gh4nv/?view_only=ddc941dfacec456482984254792bdade.

**Table 1**

*Descriptive Statistics of Hit Songs on TikTok and Spotify*

| Variables | TikTok ($N$ = 321) | | Spotify ($N$ = 1,707) | |
|---|---|---|---|---|
| | Percentage | | Percentage | |
| Major label (yes=1) | 55% | | 76% | |
| Themes[a] | | | | |
| Relationship/Love/Passion/Sexual Desire | 60% | | 63% | |
| Good Times/Partying/Carnival/Festival | 6% | | 9% | |
| Identity/Depression/Mood Issues/Spirituality | 18% | | 28% | |
| Social Politics/Anger/Violence/Race/Religion | 17% | | 19% | |
| Others | 14% | | 4% | |
| Genres[b] | | | | |
| Pop | 42% | | 62% | |
| Rock | 7% | | 7% | |
| R&B/Soul | 15% | | 9% | |
| Hip-hop/Rap/Trap | 36% | | 42% | |
| Dance | 17% | | 14% | |
| Others | 19% | | 16% | |
| | *M* | *SD* | *M* | *SD* |
| On_chart_days | 208.816 | 253.951 | 42.531 | 84.390 |
| Control variables | | | | |
| Duration (in minutes) | 3.156 | 1.030 | 3.359 | .917 |
| Peak daily popularity | 50.393 | 29.616 | 55.325 | 29.909 |
| Singer popularity | 3077529.495 | 9694120.27 | 23798444.89 | 28092554.5 |

*Note.* [a] One song can have more than one theme.
[b] One song can be categorized into more than one genre.

**Table 2**

*Mixed-Effects Linear Regression Predicting Hit Songs Popularity on TikTok and Spotify*

| Variables | TikTok ($N$ = 321) | Spotify ($N$ = 1,707) |
|---|---|---|
| Fixed effects | | |
| Independent variables | | |
| Major label | -.628 (.234) ** | -.221 (.079) ** |
| Themes | | |
| Relationship/Love/Passion/Sexual Desire | -.250 (.248) | .342 (.085) *** |
| Good Times/Partying/Carnival/Festival | .530 (.483) | -.108 (.136) |
| Identity/Depression/Mood Issues/Spirituality | -.068 (.288) | -.042 (.083) |
| Social Politics/Anger/Violence/Race/Religion | -.014 (.294) | -.372 (.102) *** |
| Genres | | |
| Pop | .195 (.232) | -.141 (.073) |
| Rock | .223 (.419) | .108 (.130) |
| R&B/Soul | .052 (.307) | -.072 (.114) |
| Hip-hop/Rap/Trap | .242 (.236) | -.104 (.072) |
| Dance | .726 (.301) * | .147 (.102) |
| Control variables | | |
| Peak daily popularity | .054 (.004) *** | .039 (.001) *** |
| Duration | .064 (.105) | -.102 (.037) ** |
| Singer popularity (log) | -.011 (.017) | -.003 (.006) |
| $\sigma^2$ | 3.613 | 1.229 |
| $\tau_{00}$ | .000 | .596 |
| AIC | 1369.093 | 5939.897 |

*Note.* * $p$ < .05; ** $p$ < .01, *** $p$ < .001. Group ID is included as a random effect.

**Figure 1**

*Daily Top-100 Hit Songs on TikTok and Spotify by Theme and Genre*

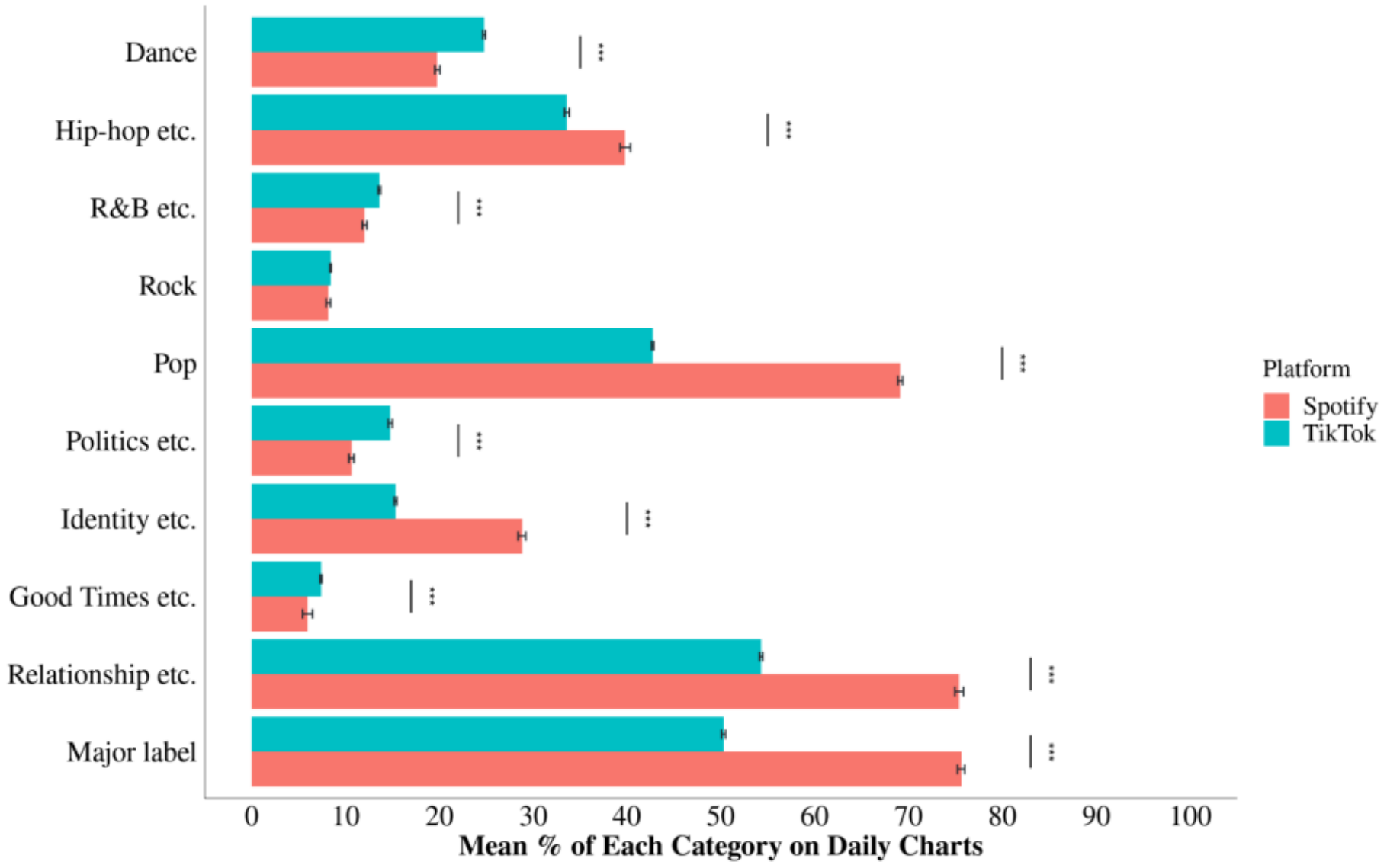

*Note*. *** *p* < .001. Error bars represent 95% confidence intervals.

Hip-hop etc.= Hip-hop/Rap/Trap, R&B etc.= R&B/Soul, Politics etc.= Social/Politics/Violence/Race/Religion, Identity etc.= Identity/Depression/Mood Issues/Spirituality, Good Times etc.= Festival/Partying/Good times, Relationship etc.=Love/Passion/Relationship/Desire.

**Figure 2**

*Hit Songs' Appearance in Top-100 Charts of TikTok and Spotify (N = 38).*

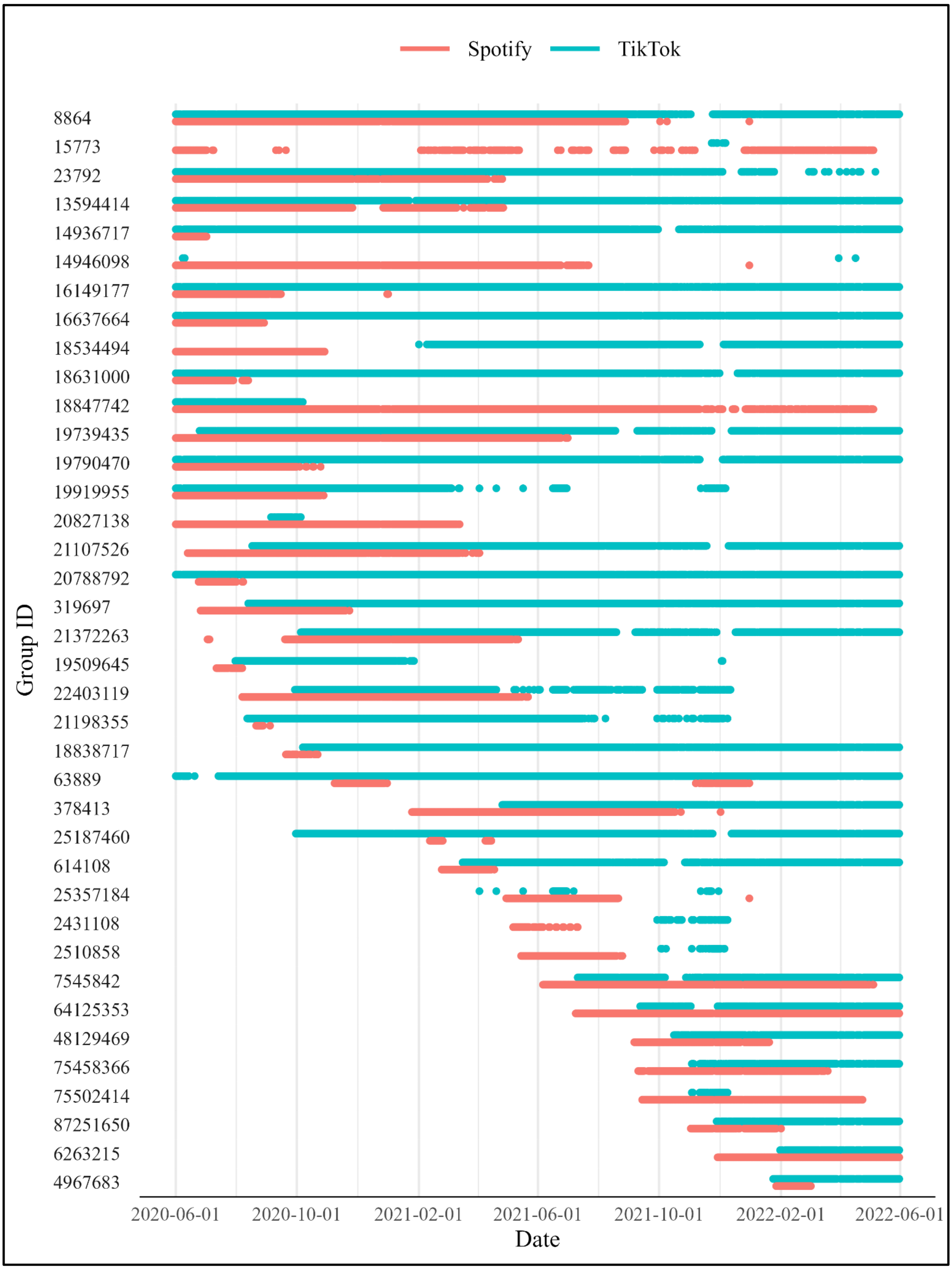

*Note.* Hit songs were included if they appeared at least twice in Top-100 lists on both platforms (*N*=38). We sorted the songs from top to bottom by the date they first appeared on Spotify charts. Songs that are cover versions of the original are assigned the same Group ID. The mapping between Group ID and song names, as well as a full visualization of appearance for songs that appeared at least once on both platforms (*N*=68) can be found in Appendix F.

# Supplementary Materials

## Appendix A. Processing of Incomplete and Erroneous Top-100 Hit Song Charts & Statistics of Themes

Due to missing data on TikTok (July 12, 2020, and April 23 & 24, 2022) and Spotify (January 4, 2021, November 25 & 26, 2021, and April 2, 2022), and erroneous data on TikTok (June 7, 2020, January 7, March 30, April 6 & 16, 2022), we removed these dates to ensure the quality of analysis. Within the 2-year time frame of our dataset, there were 100 days on TikTok where the Top-100 hit songs returned by the Chartmetric API were incomplete. Table S1 shows the details. Among these dates, January 9, 2022, was removed from the dataset because there were as many as 89 missing songs on this day. For the remaining dates, we supplemented them with the songs ranked 101-105 as needed.

The peak ranks of songs on dates 2020-06-07, 2022-01-07, 2022-01-09, 2022-03-30, 2022-04-06, and 2022-04-16 are nearly identical to their single-day rankings on these days. However, according to the trends exhibited by songs on other dates, their peak ranks could be higher, which we consider as erroneous data rather than a representation of reality. We then deleted these dates from the dataset.

Table S2 lists the theme categories and description based on lyrics.

## Table S1

*Incomplete and Erroneous Top-100 Hit Song Charts on TikTok and Data Cleaning*

| Types of error | Error dates [a] | | | Ways of cleaning/processing |
|---|---|---|---|---|
| Missing a specific rank in the top-100 charts | 2020-06-01 (89) | 2021-06-20 (67) | 2021-07-24 (90) | Supplement the missing ranks with the songs ranked 101-105 as needed |
| | 2020-06-02 (83) | 2021-06-21 (69) | 2021-07-25 (88) | |
| | 2020-06-03 (77) | 2021-06-22 (66) | 2021-07-26 (87) | |
| | 2020-06-04 (93) | 2021-06-23 (68) | 2021-07-27 (90) | |
| | 2020-06-05 (93) | 2021-06-24 (66) | 2021-07-28 (91) | |
| | 2020-06-06 (100) | 2021-06-25 (65) | 2021-07-29 (93) | |
| | 2020-06-08 (79) | 2021-06-26 (68) | 2021-07-30 (92) | |
| | 2020-06-10 (80) | 2021-06-27 (69) | 2021-07-31 (93) | |
| | 2020-06-20 (99) | 2021-06-28 (70) | 2021-08-01 (93) | |
| | 2020-12-14 (18) | 2021-06-29 (69) | 2021-08-02 (94) | |
| | 2020-12-18 (1) | 2021-06-30 (70) | 2021-08-03 (95) | |
| | 2020-12-27 (15) | 2021-07-01 (79) | 2021-08-04 (93) | |
| | 2021-01-05 (13) | 2021-07-02 (75) | 2021-08-05 (94) | |
| | 2021-05-30 (96) | 2021-07-03 (79) | 2021-08-06 (94) | |
| | 2021-05-31 (95) | 2021-07-04 (80) | 2021-08-07 (94) | |
| | 2021-06-01 (93) | 2021-07-05 (80) | 2021-08-08 (92) | |
| | 2021-06-02 (92) | 2021-07-06 (80) | 2021-08-09 (97) | |
| | 2021-06-03 (91) | 2021-07-07 (69) | 2021-08-10 (96) | |
| | 2021-06-04 (86) | 2021-07-08 (80) | 2021-08-11 (97) | |
| | 2021-06-05 (89) | 2021-07-09 (83) | 2021-08-12 (95) | |
| | 2021-06-06 (83) | 2021-07-10 (82) | 2021-08-13 (95) | |
| | 2021-06-07 (84) | 2021-07-11 (84) | 2021-08-15 (98) | |
| | 2021-06-08 (83) | 2021-07-12 (89) | 2021-08-17 (100) | |
| | 2021-06-09 (80) | 2021-07-13 (87) | 2021-08-18 (100) | |
| | 2021-06-10 (83) | 2021-07-14 (85) | 2021-08-19 (100) | |
| | 2021-06-11 (82) | 2021-07-15 (86) | 2021-08-28 (100) | |
| | 2021-06-12 (83) | 2021-07-16 (87) | 2021-08-30 (99) | |
| | 2021-06-13 (79) | 2021-07-17 (88) | 2021-10-03 (99) | |
| | 2021-06-14 (79) | 2021-07-18 (87) | 2021-11-18 (19) | |
| | 2021-06-15 (79) | 2021-07-19 (89) | 2022-01-09 (12-100) [b] | |
| | 2021-06-16 (67) | 2021-07-20 (89) | 2022-03-31 (17, 41, 50, 66, 92) | |
| | 2021-06-17 (68) | 2021-07-21 (89) | 2022-04-17 (17, 45, 60, 78) | |
| | 2021-06-18 (67) | 2021-07-22 (89) | | |
| | 2021-06-19 (69) | 2021-07-23 (89) | | |
| The peak ranks of songs on these dates are nearly identical to their rankings on the charts on that day, which is unreasonably different from the trends exhibited by songs on other dates, which we consider as due to a data error rather than a representation of reality | 2020-06-07, 2022-01-07, 2022-03-30, 2022-04-06, 2022-04-16 | | | Delete from dataset |

*Note:* [a]The numbers in brackets represent the specific ranks that Chartmetric API did not return. [b] This date was deleted rather than supplemented because too many songs are missing (up to 89 songs).

**Table S2**

*Theme Categories and Description Based on Lyrics*

| Theme | Description | Cohen's kappa |
|---|---|---|
| 1=Love/Passion/ Relationship/Desire | lovelorn experience, struggle and entanglement, passion and desire related to love and intimate relationships. | 0.878 |
| 2=Festival/Partying/ Good times | joyfulness associated with festive celebrations, gatherings of relatives and friends, blessings and prayers. | 0.945 |
| 3=Identity/Depression/ Mood Issues/Spirituality | personal confusion, sadness and depression, and problems related to negative emotions. | 0.918 |
| 4=Social/Politics/ Violence/Race/Religion | topics related to social issues, political issues, violence and rebellion. | 0.920 |
| 5=Others | other themes with no relation to the above four themes. | 0.858 |

## Appendix B. Results of Paired Samples *t*-tests Analysis to Compares the Percentage of Daily Top-100 Hit Songs between TikTok and Spotify on Each Category

To answer RQ1b, RQ2b, and RQ3b, we conducted both independent samples *t*-tests and paired samples *t*-tests. We report the results of independent samples *t*-tests in the main text ($N$=721 on TikTok and $N$ = 726 on Spotify) and present the results of paired samples *t*-tests here ($N$ = 717 on both platforms) in Table S3. Results from both tests are consistent.

**Table S3**

*Comparison of the Percentage of Daily Top-100 Hit Songs Between TikTok and Spotify on Each Category by Paired Samples t-tests Method*

| | | | Paired *t*-test | | |
|---|---|---|---|---|---|
| | *M* | *SD* | *t* value | *df* | Sig. (two-tailed) |
| Major label - TikTok | 50.277 | 2.859 | 139.15 | 716 | .000 |
| Major label - Spotify | 75.602 | 5.339 | | | |
| Relationship/Love/Passion/Sexual Desire - TikTok | 54.289 | 2.353 | -96.633 | 716 | .000 |
| Relationship/Love/Passion/Sexual Desire - Spotify | 75.387 | 6.251 | | | |
| Good Times/Partying/Carnival/Festival - TikTok | 7.380 | 1.288 | 5.031 | 716 | .000 |
| Good Times/Partying/Carnival/Festival - Spotify | 5.963 | 7.299 | | | |
| Identity/Depression/Mood Issues/Spirituality - TikTok | 15.325 | 1.971 | -62.767 | 716 | .000 |
| Identity/Depression/Mood Issues/Spirituality - Spotify | 28.798 | 5.595 | | | |
| Social Politics/Anger/Violence/Race/Religion - TikTok | 14.770 | 3.208 | 20.425 | 716 | .000 |
| Social Politics/Anger/Violence/Race/Religion - Spotify | 10.634 | 3.749 | | | |
| Pop - TikTok | 42.730 | 1.512 | -185.39 | 716 | .000 |
| Pop - Spotify | 69.118 | 3.531 | | | |
| Rock - TikTok | 8.416 | .976 | 2.052 | 716 | .041 |
| Rock - Spotify | 8.176 | 3.342 | | | |
| R&B/Soul - TikTok | 13.605 | 1.821 | 11.197 | 716 | .000 |
| R&B/Soul - Spotify | 12.030 | 3.330 | | | |
| Hip-hop/Rap/Trap - TikTok | 33.588 | 3.478 | -20.75 | 716 | .000 |
| Hip-hop/Rap/Trap - Spotify | 39.809 | 7.522 | | | |
| Dance - TikTok | 24.782 | 2.039 | 31.427 | 716 | .000 |
| Dance - Spotify | 19.792 | 3.659 | | | |

## Appendix C. Construction of Time Series of Hit Songs' Average Weekly Rankings

To identify cross platform influences of music popularity, we constructed weekly time series for hit songs using daily rankings on both TikTok and Spotify with the following steps: First, we extracted the songs included by both TikTok and Spotify's Top-100 charts based on the Group_Id, 68 song groups were obtained.

Second, we excluded the songs that were listed in the charts for less than 2 days on both platforms since it is difficult to detect the trend if a song's ranking data is too sparse.

And 38 songs were left. We also excluded the songs whose chart dates on both platforms did not overlap at all, which left us with 30 song groups.

Third, for these song groups, we got the popularity for each hit song on each platform by subtracting the original ranking of everyday from 101. Thus, a bigger popularity value means a higher ranking in the charts. Also we added up the popularity values of different song versions with the same group_id on each platform to get the final popularity of each hit song group.

**Figure S1**

*Histogram of average intervals of the number of days between exiting and reappearing in the Top-100 charts of TikTok and Spotify)*

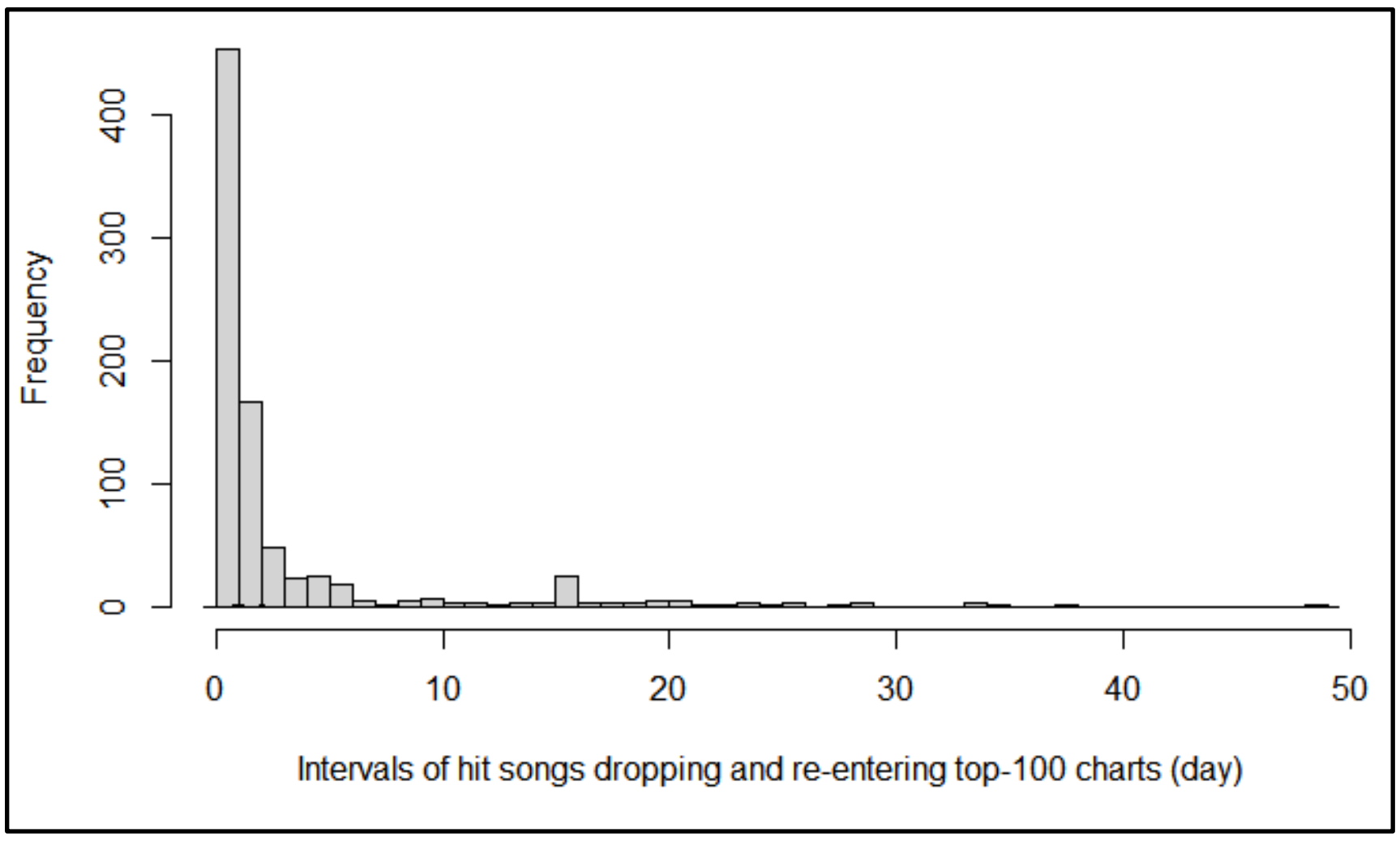

Fourth, we used data imputation forecasting techniques (Li et al., 2015) for imputation of values on those aforementioned missing dates since the further time series analyses required that there should be no missing data. Then, to mitigate the random fluctuation of peak daily popularity and derive a more stabilized time series representing the popularity trend, we used a simple moving average over a window of past *k* days as a temporal smoothing technique. Smoothing is a critical issue and the key step is to find the best window size. A window that is too small can not reduce the noise and sudden transition. Too much smoothing (a big window size), however, might deviate from the real value to some extent (Borghi et al., 2021; O'Connor et al., 2010). In the context of the songs' ranking time series, we manually calculated the intervals between each song existing and reappearing in the Top-100 charts, and plotted the histogram (as shown in Figure S1). Since the mean and median intervals are 3.259 and 1.0 respectively, and the window size is usually an odd integer (Hyndman & Athanasopoulos, 2018), we chose $k = 3$ as our smoothing time window and agreed it's suitable for reducing sudden transition and maintaining the ranking changes. Results for $k = 5$ were also obtained for robustness check and included in Appendix G.

Next, considering that the rankings of those dates that a hit song not charted were all zero which could potentially influence the interaction effects across platforms, we extracted a subset for each hit song time series based on this rule: selecting the dates on which a hit song started to enter the daily charts of any platform until it exited the charts on all platforms.

Finally, we calculated the average popularity rankings on a weekly basis for each hit song. Following the above steps, we obtained the stabilized time series of weekingly rankings for the selected 30 hit songs.

**Appendix D. The Analysis Processes of VAR models**

For VAR analyses, all time series must be stationary. We in the first step conducted *Augmented Dickey-Fuller* (ADF) tests to ensure the time series did not contain unit root. The ADF test that accounts for the complicated structure of most time series is used to test whether a series is stationary. If the ADF test indicates that the series is non-stationary, the series is said to contain a 'unit root' (For details, see, Vliegenthart & Montes, 2014). All time series for TikTok and Spotify were stationary after differencing them. Next, we decided the appropriate number of lags for each VAR model. The optimal lag means how many days are needed for the mutual influence of two variables in the model to occur. A set of statistics can be used to assess the lags, such as the Akaike's information criterion (AIC), the Hannan and Quinn information criterion (HQIC), the final prediction error (FPE), and the Schwarz's Bayesian information criterion (SBIC). In this paper, a maximum of 4 weeks had been tested for each VAR model, and AIC was utilized to determine the optimal lag drawing on previous study (Liew, 2004). According to Brandt and Williams (2007), the number of lags needed to be limited in order to prevent using too many degrees of freedom which can result in inefficient estimates. When the given optimal lags were different, the relatively small lag was selected.

After estimating the VAR model, Granger causality was calculated to evaluate whether the songs' dynamic popularity on Spotify (e.g, the ranking trend) is associated with the dynamic popularity on Tiktok, or vice versa. A variable *y* is assumed to Granger cause a variable *z* if the prediction of *z* based on its own past values improves after adding *y*'s past to the equation (Brandt & Williams, 2007). We then applied the cumulative impulse response functions (CIRF) and the forecast error vector decomposition (FEVD) to assess the direction and the size of the influences emerging from the Granger causality tests (Becketti, 2013; Jansen et al., 2018). Specifically, the Impulse response functions (IRFs) from the VAR models provide further information regarding the longer-term effects of one variable (X) on another (Y), by testing the impact or "shock" of X on Y (Swanson & Granger, 1997). IRFs are particularly useful for determining the statistical significance, magnitude, and temporal pattern of one variable to a one standard deviation increase in another variable, while controlling for other variables in the model (Zhang et al., 2021). Also, a presentation of the decomposition of the Forecast Error Variance (FEVD), indicating for each variable over-time what portion of the difference between the actual and predicted values can be attributed to each of the variables in the equation. In other words, this method estimates over-time the amount of variation in each of the endogenous variables that can be attributed to its own past and to the past of each of the other endogenous variables (Vliegenthart, 2014).

**Appendix E. Robustness Checks of Temporal Order of Hit Songs' Appearance on Two Platforms**

Considering the influences of the chosen time period, we conducted the robustness checks to examine the temporal order of hit songs' appearance on two platforms. First, we excluded the songs that entered the daily charts of either platform on the first day of our observing time window (June 1st, 2020) and 41 songs were left. Chi-square tests were conducted and results revealed that for all 41 songs listed on both platforms, 36(87.8%) of them entered the Spotify daily charts earlier than TikTok charts, while 5 (12.2%) of them firstly entered the TikTok daily charts ($\chi^2(1, N = 41) = 32.193$, $p < 0.001$). Second, we excluded the songs that dropped off the daily charts of either platform on the last day of our observing time window (May 31st, 2022) and 40 songs were left. Meanwhile, 32 of the 40 hit songs (80%) dropped off the Spotify daily charts earlier than TikTok daily charts while 8 of them first dropped off the TikTok daily charts ($\chi^2(1, N = 40) = 33.994$, $p < 0.001$). These results supported our previous findings.

**Appendix F. 68 Hit Songs' Appearance in Top-100 Charts of Spotify and TikTok**

There were 68 hit songs that appeared at least once on both platforms, with or without overlapping day(s), and 38 songs that appeared at least twice on both platforms. Table S4 shows the mapping between Group ID and song names on two platforms. Figure S2 shows the temporal order of all 68 hit songs' appearances in the Top-100 charts of Spotify and TikTok.

**Table S4**

*The Mapping Between Group ID and Song Names on Spotify and TikTok*

| ID | Group ID | TikTok song names (with different versions) | Spotify song names (with different versions) |
|---|---|---|---|
| 1 | 5030 | Sweater-Weather-222332331146174464 | Sweater Weather |
| 2 | 7900 | Spooky-Scary-Skeletons-6602666381370460933 | Spooky, Scary Skeletons - Undead Tombstone Remix |
| 3 | 8864 | Dance-Monkey-6717552289336314629 | Dance Monkey |
| 4 | 15773 | Believer-249539203825504256 | Believer |
| 5 | 18673 | Boss-Bitch-6787078702082607878 | Boss Bitch |
| 6 | 23792 | death-bed-coffee-for-your-head-6703427142346083074 | death bed (coffee for your head) |
| 7 | 25259 | Favorito-6807475089290823682 | Favorito |
| 8 | 60807 | Stunnin' (feat. Harm Franklin) | Stunnin-feat-Harm-Franklin-6813134956269947654 |
| 9 | 63889 | All-I-Want-for-Christmas-is-YOU-172594401737605120, All-I-Want-for-Christmas-Is-You-222454852260487168 | All I Want for Christmas Is You |
| 10 | 90172 | Heartbreak-Anniversary-6794920067080390657 | Heartbreak Anniversary |
| 11 | 319697 | How-You-Like-That-6842582526297393154 | How You Like That |
| 12 | 348355 | Deep-End-6854654862202784517 | Deep End |
| 13 | 353158 | Haw®¢i - Remix | Haw®¢i |
| 14 | 353529 | Dynamite-6862966224016377857 | Dynamite, Dynamite - Instrumental |
| 15 | 376133 | Iko-Iko-My-Bestie-6754595656993605633 | Iko Iko (My Bestie) |
| 16 | 378413 | Astronaut-In-The-Ocean-6816496693551450885, Astronaut-In-The-Ocean-6914530565300226049 | Astronaut In The Ocean |
| 17 | 614108 | SugarCrash-6920125567752734722 | SugarCrash! |
| 18 | 1047351 | Calling-My-Phone-6907410938765413126 | Calling My Phone |
| 19 | 2431108 | Bipolar-6948841318537841413 | Bipolar |
| 20 | 2510858 | Build-a-Btch-6956990112127585029 | Build a Bitch |
| 21 | 3344905 | double-take-6976066440176749313 | double take |
| 22 | 3691682 | J-Cole-No-Role-Modelz-272907332983189504 | No Role Modelz |
| 23 | 4967683 | Ginseng-Strip-2002-6997999797663714054 | Ginseng Strip 2002 |
| 24 | 6263215 | Infinity-6704992801559414786 | Infinity |
| 25 | 6537179 | DARARI-7065014354940349210 | DARARI |
| 26 | 7545842 | Beggin'-5000000001320781379 | Beggin' |
| 27 | 10159602 | It's-Beginning-to-Look-a-Lot-Like-Christmas-Perry-Como-The | It's Beginning to Look a Lot like Christmas |
| 28 | 13594414 | Se?orita-6704854531001289474 | Se?orita |
| 29 | 14936717 | Lose-Control-6743772128031557633 | Lose Control |
| 30 | 14946098 | Roses-Imanbek-Remix-6738439639826287365 | Roses - Imanbek Remix |
| 31 | 16149177 | Say-So-6763054442704145158, Say-So-by-Doja-Cat-6778164370233953029 | Say So, Say So (feat. Nicki Minaj) |
| 32 | 16151184 | Streets-6909978363763526405 | Streets |
| 33 | 16637664 | Supalonely-feat-Gus-Dapperton-6759409576673560577 | Supalonely |
| 34 | 17303377 | This-City-6685871967402199814 | This City |
| 35 | 17340502 | Watermelon-Sugar-6769291961120589825 | Watermelon Sugar |
| 36 | 18383653 | Yummy-6777509948566800385 | Yummy |
| 37 | 18534494 | ily-i-love-you-baby-6798329661525854210 | ily (i love you baby) (feat. Emilee) |
| 38 | 18631000 | Savage-6800996740322297858, Savage-Tiger-King-Edition-6812049289192196870 | Savage, Savage (feat. Beyoncé) - Remix |
| 39 | 18708225 | After-Party-6803571286871115778 | After Party |
| 40 | 18837299 | Skechers-6788456091559676678 | Skechers |
| 41 | 18838717 | You-Got-It-6796126109965502465 | You Got It |
| 42 | 18847742 | Don't-Start-Now-6782222148476947205 | Don't Start Now |
| 43 | 19509645 | Rags2Riches-6814905955071953670 | Rags2Riches (feat. ATR Son Son) |
| 44 | 19617946 | MAMACITA-6813015571236390913 | MAMACITA |
| 45 | 19732147 | you-broke-me-first-6818936887760259846 | you broke me first |
| 46 | 19739435 | ROCKSTAR-6818447895918807041 | ROCKSTAR (feat. Roddy Ricch), ROCKSTAR (feat. Roddy Ricch) - BLM REMIX |
| 47 | 19790470 | Party-Girl-6802011143096339205 | Party Girl, Party Girl (Remix) |
| 48 | 19831222 | THE-SCOTTS-6819262113299565318 | THE SCOTTS |
| 49 | 19919955 | Toosie-Slide-6811422424131700738 | Toosie Slide |
| 50 | 20788792 | Banana-feat-Shaggy-6788784989656926981 | Banana (feat. Shaggy) - DJ FLe - Minisiren Remix |
| 51 | 20827138 | Relación-6812538887328532481, Relationship-Slow-Version-6785167255597566725 | Relación |
| 52 | 20939448 | TKN-6831569851157317633 | TKN (feat. Travis Scott) |
| 53 | 21107526 | Savage-Love-Laxed-Siren-Beat-6825494114277100293, Savage-Love-Laxed-Siren-Beat-6828290428744338181, Savage-Love-Laxed-Siren-Beat-BTS-Remix-6878409032286095362 | Savage Love (Laxed - Siren Beat), Savage Love (Laxed – Siren Beat) [BTS Remix] |
| 54 | 21198355 | Tap-In-6839497849235950341 | Tap In (feat. Post Malone, DaBaby & Jack Harlow), Tap In (feat. Post Malone, DaBaby & Jack Harlow) |
| 55 | 21372263 | What-You-Know-Bout-Love-6847647314127932166 | What You Know Bout Love |
| 56 | 21812796 | Mood-feat-iann-dior-6851608429664929794 | Mood (feat. iann dior), Mood (Remix) feat. |
| | | | Justin Bieber, J Balvin & iann dior |
| 57 | 22403119 | WAP-Megan-Thee-Stallion-6858616259425225478, WAP-Megan-Thee-Stallion-6858456364713282309 | WAP (feat. Megan Thee Stallion) |
| 58 | 23173861 | Put-Your-Records-On-6815868828203878402 | Put Your Records On |
| 59 | 23873313 | WITHOUT-YOU-6892161420423596034 | WITHOUT YOU, WITHOUT YOU (with Miley Cyrus) |
| 60 | 25187460 | Love-Story-6842883579551255302 | Love Story (Taylor's Version) |
| 61 | 25357184 | Rasputin-Single-Version-242977707800637440 | Rasputin |
| 62 | 40743625 | His-Hers-6964112488388397830 | His & Hers (feat. Don Toliver, Lil Uzi Vert & Gunna) |
| 63 | 48129469 | Life-Goes-On-7000772466247289606, Life-Goes-On-7000772466247289606 | Life Goes On |
| 64 | 64125353 | STAY-6981869640796342274 | STAY (with Justin Bieber) |
| 65 | 75458366 | MONEY-7005888397580208130 | MONEY |
| 66 | 75502414 | Love-Nwantiti-Dance-Ver-7003020102824512262 | Love Nwantiti - Remix, love nwantiti (ah ah ah), love nwantiti (feat. ElGrande Toto) - North African Remix |
| 67 | 77003351 | MONTERO-Call-Me-By-Your-Name-6942551356402042881 | MONTERO (Call Me By Your Name) |
| 68 | 87251650 | SAD-GIRLZ-LUV-MONEY-Remix-7007785041036101634 | SAD GIRLZ LUV MONEY Remix (feat. Kali Uchis and Moliy) |

**Figure S2**

*Hit Songs' Appearance in Top-100 Charts of TikTok and Spotify (N = 68).*

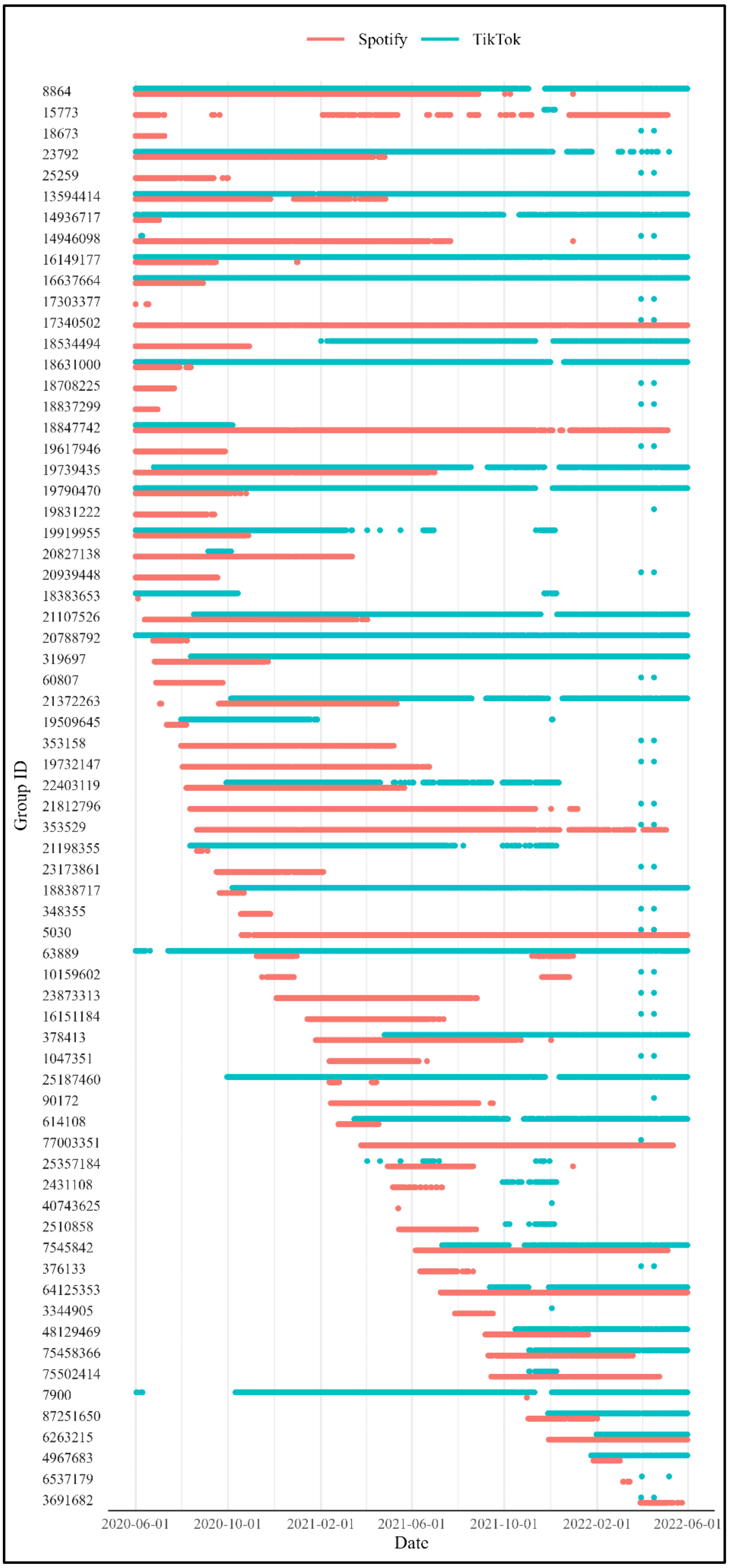

*Note.* Songs are sorted from top to bottom by the date they first appeared on Spotify hit song charts.

## Appendix G. Results of VAR Models by Using k = 3 and k = 5 as the Smoothing Windows

Table S5 shows the results of VAR models predicting the popularity of 30 hit songs on TikTok and Spotify with the smoothing window k = 3.

**Table S5**

*Results of VAR Models Predicting the Popularity of 30 Hit Songs on TikTok and Spotify*

| | Spotify Predicting TikTok | | | | TikTok Predicting Spotify | | | |
|---|---|---|---|---|---|---|---|---|
| Song name | *F* | *p*-value | FEVD | CIRF | *F* | *p*-value | FEVD | CIRF |
| Dance Monkey | 0.030 | 0.993 | | | 0.030 | 0.993 | | |
| Death Bed (coffee for your head) | 0.000 | 0.983 | | | 0.039 | 0.884 | | |
| All I Want for Christmas Is You | 2.798 | 0.027* | 0.083 | -1.710 | **3.018** | **0.019*** | **0.107** | **-7.208** |
| How You Like That | 0.181 | 0.671 | | | 0.062 | 0.803 | | |
| Astronaut In The Ocean | 0.023 | 0.880 | | | 0.005 | 0.946 | | |
| SugarCrash | 1.185 | 0.322 | | | **2.594** | **0.040*** | **0.181** | **-2.246** |
| Ginseng Strip 2002 | 0.959 | 0.436 | | | 0.023 | 0.995 | | |
| Infinity | 0.060 | 0.808 | | | 0.101 | 0.752 | | |
| Beggin' | 0.017 | 0.896 | | | 0.382 | 0.538 | | |
| Señorita | 2.983 | 0.020* | 0.017 | -0.772 | 0.798 | 0.528 | | |
| Lose Control | 0.009 | 1.000 | | | 0.012 | 1.000 | | |
| Say So | 0.008 | 1.000 | | | 0.009 | 1.000 | | |
| Supalonely | 0.082 | 0.922 | | | 0.004 | 0.996 | | |
| Savage | 1.964 | 0.102 | | | 0.258 | 0.905 | | |
| You Got It | 6.417 | 0.000*** | 0.009 | -0.529 | 1.704 | 0.152 | | |
| Don't Start Now | 0.363 | 0.780 | | | 0.564 | 0.639 | | |
| Rags2Riches | 6.912 | 0.000*** | 0.034 | 0.134 | 33.044 | 0.000*** | 0.256 | -0.063 |
| Rockstar | 0.003 | 0.959 | | | 0.001 | 0.977 | | |
| Party Girl | 0.010 | 0.999 | | | 0.308 | 0.819 | | |
| Toosie Slide | 0.926 | 0.451 | | | 2.606 | 0.039* | 0.073 | -0.492 |
| Relación | 2.416 | 0.075 | | | 0.975 | 0.411 | | |
| Savage Love (Laxed - Siren Beat) | 0.098 | 0.755 | | | 0.006 | 0.941 | | |
| What You Know Bout Love | 2.784 | 0.064 | | | 0.007 | 0.993 | | |
| Love Story (Taylor's Version) | 0.040 | 0.989 | | | 0.006 | 0.999 | | |
| Rasputin | 0.302 | 0.584 | | | 0.032 | 0.859 | | |
| Life Goes On | 4.526 | 0.006** | 0.261 | 3.719 | **4.934** | **0.004**** | **0.131** | **-5.576** |
| Stay (with Justin Bieber) | 1.165 | 0.284 | | | 0.002 | 0.968 | | |
| Money | 0.053 | 0.818 | | | 0.077 | 0.782 | | |
| Love Nwantiti | 0.331 | 0.803 | | | **15.285** | **0.000***** | **0.055** | **-12.426** |
| Sad Girlz Luv Money | 4.688 | 0.004** | 0.053 | 1.613 | **6.812** | **0.000***** | **0.409** | **-5.034** |

*Note*. Significance codes for Granger causality tests: *$p < .05$. **$p < .01$. ***$p < .001$. FEVD = forecast error variance decomposition, CIRF = cumulative impulse response functions. The optimal lag for each song group (each model) was chosen based on Akaike Information Criterion (AIC). CIRF and FEVD for n-lag steps are only shown for significant Granger causality results. Significant CIRF results based on 95% bootstrap confidence intervals (i.e., upper and lower intervals do not cross 0) are in bold.

As explained in Appendix C, the smoothing window k = 5 was also employed for robustness check. Following the same steps, we obtained the weekly time series for all hit songs. The VAR models and Granger Causality tests were also conducted and results are shown in Table S6. Similar significant results were found: CIRF results showed none of the hit song's popularity on Spotify predicted the subsequent TikTok popularity and TikTok popularity negatively predicted Spotify popularity significantly for the same five songs (*All I Want for Christmas Is You*, *SugarCrash*, *Love Nwantiti*, *Life Goes On* and *Sad Girlz Luv Money*).

## Table S6

*Results of VAR Models Predicting the Popularity of 30 Hit Songs on TikTok and Spotify with the Smoothing Window K = 5*

| | Spotify Predicting TikTok | | | | TikTok Predicting Spotify | | | |
|---|---|---|---|---|---|---|---|---|
| Song name | *F* | *p*-value | FEVD | CIRF | *F* | *p*-value | FEVD | CIRF |
| Dance Monkey | 0.023 | 0.995 | | | 0.024 | 0.995 | | |
| Death Bed (coffee for your head) | 0.015 | 0.902 | | | 0.080 | 0.777 | | |
| All I Want for Christmas Is You | 3.071 | 0.018* | 0.118 | -2.788 | **2.124** | **0.080†** | **0.084** | **-5.841** |
| How You Like That | 1.541 | 0.206 | | | 0.362 | 0.780 | | |
| Astronaut In The Ocean | 0.036 | 0.850 | | | 0.019 | 0.890 | | |
| SugarCrash | 1.523 | 0.201 | | | **2.461** | **0.050†** | **0.182** | **-2.137** |
| Ginseng Strip 2002 | 0.619 | 0.659 | | | 0.224 | 0.919 | | |
| Infinity | 0.032 | 0.858 | | | 0.015 | 0.901 | | |
| Beggin' | 0.000 | 0.992 | | | 0.556 | 0.458 | | |
| Señorita | 1.046 | 0.353 | | | 0.023 | 0.977 | | |
| Lose Control | 0.043 | 0.996 | | | 0.041 | 0.997 | | |
| Say So | 0.008 | 1.000 | | | 0.009 | 1.000 | | |
| Supalonely | 0.109 | 0.897 | | | 0.002 | 0.998 | | |
| Savage | 0.825 | 0.511 | | | 0.254 | 0.907 | | |
| You Got It | 6.678 | 0.000*** | 0.009 | -0.505 | 1.546 | 0.192 | 0.021 | 0.041 |
| Don't Start Now | 0.197 | 0.939 | | | 0.479 | 0.751 | | |
| Rags2Riches | 7.177 | 0.000*** | 0.043 | -0.233 | 40.558 | 0.000*** | 0.239 | -0.006 |
| Rockstar | 0.095 | 0.759 | | | 0.011 | 0.915 | | |
| Party Girl | 0.009 | 0.999 | | | 0.385 | 0.764 | | |
| Toosie Slide | 0.137 | 0.938 | | | 3.637 | 0.014* | 0.068 | -0.362 |
| Relación | 2.492 | 0.068 | | | 2.693 | 0.054 | | |
| Savage Love (Laxed - Siren Beat) | 0.087 | 0.967 | | | 0.003 | 1.000 | | |
| What You Know Bout Love | 2.863 | 0.060 | | | 0.008 | 0.992 | | |
| Love Story (Taylor's Version) | 0.058 | 0.982 | | | 0.004 | 1.000 | | |
| Rasputin | 0.264 | 0.609 | | | 0.264 | 0.609 | | |
| Life Goes On | 0.733 | 0.574 | | | **10.819** | **0.000*** ** | **0.529** | **-7.390** |
| Stay (with Justin Bieber) | 0.244 | 0.623 | | | 0.002 | 0.966 | | |
| Money | 0.003 | 0.958 | | | 0.045 | 0.832 | | |
| Love Nwantiti | 0.242 | 0.866 | | | **15.527** | **0.000*** ** | **0.056** | **-13.304** |
| Sad Girlz Luv Money | 3.792 | 0.011* | 0.041 | 2.544 | **5.825** | **0.001** ** | **0.402** | **-4.536** |

*Note*. Significance codes for Granger causality tests: † $p < .1$, *$p < .05$, **$p < .01$, ***$p < .001$. FEVD = forecast error variance decomposition, CIRF = cumulative impulse response functions. The optimal lag for each song group (each model) was chosen based on Akaike Information Criterion (AIC). CIRF and FEVD for n-lag steps are only shown for significant Granger causality results. Significant CIRF results based on 95% bootstrap confidence intervals (i.e., upper and lower intervals do not cross 0) are in bold.